\documentclass[]{aa}
\usepackage[varg]{txfonts}
\usepackage{graphicx}
\usepackage{booktabs}
\usepackage{amsmath}
\usepackage{colortbl}
\usepackage{subfig}
\usepackage{rotating}
\usepackage{amssymb}
\usepackage{latexsym}
\usepackage{ifthen}
\usepackage{enumitem}

\begin{document}

\title{An X-ray/SDSS sample (II): outflowing gas plasma properties} 

\author{M. Perna
		\inst{\ref{i1},\ref{i2}}\thanks{E-mail: michele.perna4@unibo.it}
		\and 
	G. Lanzuisi
		\inst{\ref{i1},\ref{i3}}	
		\and
	M. Brusa
		\inst{\ref{i1},\ref{i3}} 
		\and
	G. Cresci
		\inst{\ref{i2}}
         \and
	M. Mignoli
		\inst{\ref{i3}}
}


\institute{Dipartimento di Fisica e Astronomia, Universit\`a di Bologna, viale Berti Pichat 6/2, 40127 Bologna, Italy\label{i1}
	\and
	INAF - Osservatorio Astrofisico di Arcetri, Largo Enrico Fermi 5, 50125 Firenze, Italy\label{i2}
	\and
	INAF - Osservatorio Astronomico di Bologna, via Ranzani 1, 40127 Bologna, Italy\label{i3}
}

\date{Received 2 November 1992 / Accepted 7 January 1993}

\abstract {} 
{
Galaxy-scale outflows are nowadays observed in many active galactic nuclei (AGNs); however, their characterisation in terms of (multi-) phase nature, amount of flowing material, effects on the host galaxy, is still unsettled. In particular, ionized gas mass outflow rate and related energetics are still affected by many sources of uncertainties. In this respect, outflowing gas plasma conditions, being largely unknown, play a crucial role. 
} 
{
Taking advantage of the spectroscopic analysis results we obtained studying the X-ray/SDSS sample of 563 AGNs at z $<0.8$  presented in our companion paper, we analyse stacked spectra and sub-samples of sources with high signal-to-noise temperature- and density-sensitive emission lines to derive the plasma properties of the outflowing ionized gas component. For these sources, we also study in detail various diagnostic diagrams to infer information about outflowing gas ionization mechanisms. 
}
{
We derive, for the first time, median values for electron temperature and density of outflowing gas from medium-size samples ($\sim 30$ targets) and stacked spectra of AGNs. Evidences of shock excitation are found for outflowing gas. 
} 
{
We measure electron temperatures of the order of $\sim 1.7\times10^4$ K and densities of $\sim 1200$ cm$^{-3}$ for faint and moderately luminous AGNs (intrinsic X-ray luminosity $40.5<log(L_X)<44$ in the 2-10 keV band).
We caution that the usually assumed electron density  ($N_e=100$ cm$^{-3}$) in ejected material might result in relevant overestimates of flow mass rates and energetics and, as a consequence, of the effects of AGN-driven outflows on the host galaxy. 

} 

\maketitle
\titlerunning{X-ray loudness and outflows} 

\section[Introduction]{Introduction}

\begin{figure*}[t]
\centering
\includegraphics[width=19cm,height=12cm,trim=13 0 20 0,clip,angle=0]{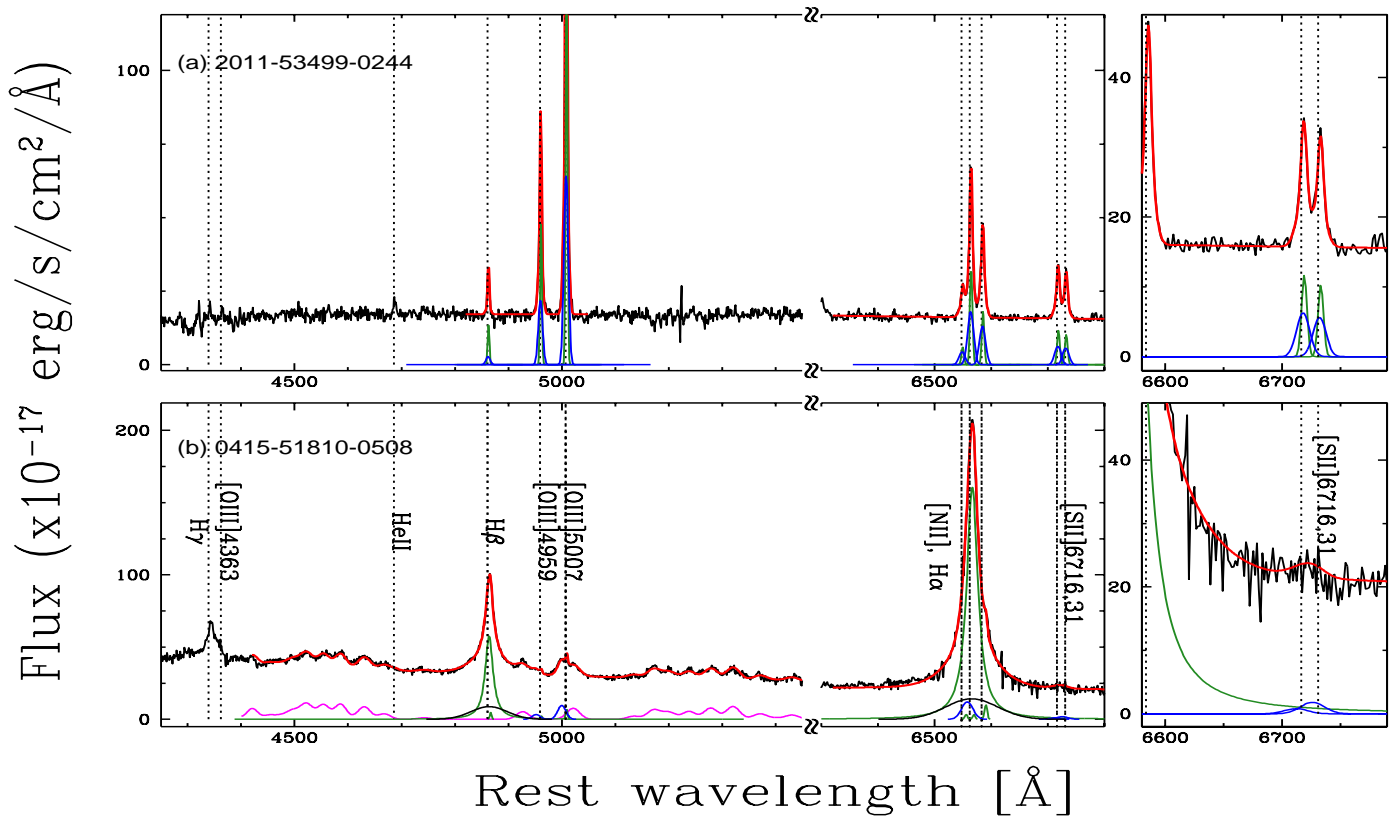} 
\caption{\small Rest-frame optical spectra of two AGNs from our X-ray/SDSS sample. For each target, in the upper left corner of the large panel we show the MJD, the plate and the fibre numbers which uniquely identify the SDSS spectrum. The dashed vertical lines mark the location of the most prominent emission lines, besides some features crucial for the analysis presented in this paper and/or in Paper I. Fit results obtained in Paper I, from the multicomponent simultaneous fit in the H$\beta$-[O {\small III}] and H$\alpha$-[NII] regions are also shown: best-fit narrow component (NC) and broad line region component (BC) profiles are highlighted with green curves; outflow component (OC) and Fe {\small II} emission are shown with blue and magenta curves, respectively. Finally,  black Gaussian profile in the low panel shows a second set of BC used to fit the complex BLR profiles (see Paper I for further details).
The insets on the right of each panel show a zoom in the vicinity of sulfur lines. 
The spectra display faintness and blending problems affecting density-sensitive [S II]$\lambda\lambda$6716,6731 and temperature-sensitive [O III]$\lambda$4363 emission lines. In particular, the object in the first panel displays well detected and modelled sulfur doublet, from which we can easily derive the density-sensitive flux ratio, but faint oxygen $\lambda$4363 line; the spectrum in the lower panel shows blended sulfur doublet features and oxygen line at $\lambda4363$. For both the AGNs it is therefore not possible to derive temperature-sensitive flux ratios. }
\label{esempi}
\end{figure*}

According to the most popular active galactic nuclei (AGN)-galaxy evolutionary models (e.g. \citealt{Hopkins2008,Menci2008,Sijacki2015}), the growth of super-massive black hole (SMBH) has a significant impact on host galaxy evolution. 
The accompanying released accretion energy, coupling with the interstellar medium (ISM), has been postulated to regulate both the star formation processes in the host and the accretion onto the SMBH (e.g. \citealt{King2015,Zubovas2014}). 

The presence of AGN-driven outflows is nowadays quite well established through high resolution observations of local and  high-redshift galaxies, 
and it is now possible to study in details the feedback phenomena, characterising the galaxy-wide extension and the morphology of the ejected material as well as the masses and the energetics related to outflows (e.g. \citealt{Bischetti2016,Brusa2016,Cresci2015, Feruglio2015,Harrison2012, Harrison2014,Husemann2016,Perna2015a,Perna2015b,Rupke2013b}; see also \citealt{Fiore2017} for an updated and complete list).
However, the physical processes responsible for the coupling between AGN winds and ISM remain largely unknown; furthermore, outflow effects on host galaxy evolution, as derived from outflow energetics (mass outflow rate, kinetic and momentum powers)  stay mostly unsettled. 

In Perna et al. (2017a; Paper I hereinafter) we presented a sample of 563 X-ray selected AGN at z $<0.8$, for which SDSS spectra are available. We combined ionized emission line and neutral absorption feature information modelled through multicomponent simultaneous fitting (\citealt{Brusa2015}), non-parametric measurement (\citealt{Zakamska2014}) and penalised pixel fitting procedure (pPXF; \citealt{Cappellari2004,Cappellari2016}) analysis, to derive kinematic properties of both warm and cold gas components of the ISM. 
The modelling of optical spectra permitted to derive the incidence of ionized ($\sim40\%$) and atomic ($< 1\%$) outflows covering a wide range of AGN bolometric luminosity, from $10^{42}$ to $10^{46}$ erg/s and to relate the presence of ionized outflows with different AGN power tracers. We also derived X-ray and bolometric luminosities and discussed our results in the context of an evolutionary sequence allowing two distinct stages for the feedback phase: an initial stage characterised by X-ray/optical obscured AGNs in which the atomic gas is still present in the ISM and the outflow processes involve all the gas components, and a later stage associated with unobscured AGNs, which line of sight has been cleaned and the cold components have been heated or exhausted. 

In this second paper we focus on the physical conditions of ionized gas, studying in details stacked spectra and small/medium-size sub-samples of X-ray/SDSS sources characterised by the presence of well detected optical diagnostic lines.
In particular, we focus on the estimate of the electron temperature ($T_e$) and density ($N_e$) of the unperturbed and outflowing ionised gas in the narrow line region (NLR). 

Plasma properties of NLR unperturbed gas are nowadays well constrained to average $N_e$ of the order of $\approx 250-400$ cm$^{-3}$ and $T_e$ of $\sim 1.5\times 10^4$ K (e.g. \citealt{Vaona2012,Zhang2013}). The physical conditions within the outflowing regions are instead mostly unknown, because of the faintness of the outflow wings of the emission lines involved in the diagnostics used to derive such information (see, e.g. \citealt{DeRobertis1986,Rice2006,Vaona2012}). The knowledge of these properties is crucial for the investigation of the mechanisms responsible of outflows: it can significantly reduce the actual uncertainties in the outflow energetics (up to a factor of ten), improving our understanding of the AGN outflow phenomenon and its impact on galaxy evolution. 

The paper is organised as follows. We first give a brief description of the spectroscopic modelling results obtained in Paper I (Section \ref{previous}). In Sect. \ref{plasma} we review the current knowledges about the NLR plasma conditions and their role in deriving the outflow energetics. In Sect. \ref{densitysection} and \ref{temperaturesection} we derive temperature- and density-sensitive flux ratios from single targets and stacked spectra. Section \ref{resultsection} displays the median plasma properties. In Sect. \ref{mechanisms} we investigate several diagnostics to infer information about outflowing gas ionization mechanisms. Finally, we summarise our results and their implications in the Discussion section (Sect. \ref{sdssdiscussion}).

\section{Previous work - optical spectroscopic analysis}\label{previous}
In Paper I we presented modelling results for a sample of 563 AGNs. Thanks to multicomponent simultaneous fit technique, for each emission line in  the wavelength range between He {\small I}$\lambda$4687 and [S {\small II}]$\lambda\lambda$6716,6731 doublet,  we were able to separate NLR gas in virial motion, modelled with narrow gaussian components [NC, which full width at half maximum (FWHM) has been constrained to be $\lesssim550$ km/s], from perturbed outflowing material, modelled with outflow components (OC; FWHM $> 550$ km/s; see, e.g. Fig. \ref{esempi} and Fig. \ref{fit}; see also Paper I, fig. 3). The outflow component detection has been tested considering chi-square minimisation and signal-to-noise (Paper I, sect. 3).  

 In this paper we take advantage of SDSS spectra modelling results to construct stacked spectra and select well defined sub-samples of sources with the intention of studying NLR plasma properties in presence of AGN-driven outflows.

\section{The plasma properties in the outflowing gas}\label{plasma}

Electron density and electron temperature within AGN-driven outflow regions are largely unknown. These quantities are nowadays important sources of uncertainties in outflow kinematic estimates for the ionised phase.
\cite{Carniani2015} showed how $N_e$ and $T_e$ enter in the determination of the ejected mass $M_{out}$ and, in consequence, of the mass outflow rate $\dot M_{out}$, kinetic  and momentum powers ($\dot E=0.5 \dot M_{out} V_{out}^2$ and $\dot P=\dot M_{out} V_{out}$, respectively; $V_{out}$ is the outflow velocity). Here we report the Eq. (5) they derived for the [O {\small III}]$\lambda$5007 line (but the same following considerations apply when Balmer emissions are used instead of [O {\small III}]; see, e.g. \citealt{Cresci2015,Liu2013}): 
\begin{equation}\label{eq1}
M_{[OIII]}= 1.7\times 10^3 \frac{m_p C L_{[OIII]}}{10^{[O/H]-[O/H]_\odot}j_{[OIII]}<N_e>}\ g ,
\end{equation}

where $m_p$ is the proton mass, $C=<N_e>^2/<N_e^2>$ is the 'condensation factor', $L_{[OIII]}$ is the [O {\small III}]$\lambda$5007 luminosity, $10^{[O/H]-[O/H]_\odot}$ is the metallicity, $j_{[OIII]}$ the oxygen emissivity. The emissivity term shows a weak dependence on electron density over several order of magnitudes, but also a relevant dependence on the electron temperature: a difference of a factor of three easily emerges when we consider $T_e$ of $1\times$ instead of $2\times 10^4$ K \footnote{[O {\small III}] emissivity from PyNeb (\citealt{Luridiana2015})}. Moreover, the outflow mass shows an inverse proportionality to the electron density. 

Outflow energetics have been usually derived in the past assuming given values for electron temperature and density. While a general consensus is found for a $T_e =1\times 10^4$ K (e.g. \citealt{Harrison2014,Carniani2015,Bischetti2016,Nesvadba2006}; but see \citealt{Liu2013,Cresci2015}), several values are used for the electron density, spanning one order of magnitude or more: for example, 1000 cm$^{-3}$ has been assumed by \cite{CanoDiaz2012}, 500 cm$^{-3}$ by \cite{Carniani2015}, and 100 cm$^{-3}$ by a large number of other authors (e.g. \citealt{Brusa2015,Cresci2015,Harrison2014,Kakkad2016,Liu2013}).

Few diagnostic ratios involving forbidden lines can be used to derive these properties in regions with densities  $\lesssim$ 10$^6$ cm$^{-3}$ (depending on the critical density of the involved forbidden transitions). 
In particular, [S {\small II}]$\lambda\lambda$6716,6731 and [O {\small III}] flux ratio (involving [OIII]$\lambda\lambda$4959,5007 and [OIII]$\lambda$4363) diagnostics, are potentially useful to measure $N_e$ and $T_e$ because of their optical wavelengths\footnote{[OII]$\lambda$3727 doublet ratio allows a further density diagnostic. However, from the observational point of view, [O {\small II}] lines are so close in wavelength that only high spectral resolution observations permit the derivation of their flux ratio. This argument precludes the use of single ionized oxygen diagnostic for the sources presented in these works.}, through the equations: 

\begin{equation}\label{eqoiii}
T_e\approx \frac{3.29\times 10^4}{ln\left ( \frac{R_{[OIII]}}{7.90}\right)}\ K,
\end{equation} 

\begin{equation}\label{eqrsii}
N_e=10^2 \sqrt{T_e} \frac{R_{[SII]} -1.49}{5.61-12.8\times R_{[SII]} }\ cm^{-3},
\end{equation}

where $R_{[O III]}=[f(\lambda5007)+f(\lambda4959)]/f(\lambda4363)$ and $R_{[S II]}=f(\lambda6716)/f(\lambda6731)$ \citep{Osterbrock2006}. 

Unfortunately, the faintness of the involved emission lines (in particular, [O {\small III}]$\lambda$4363 and [S {\small II}] doublet) and the interdependence between $T_e$ and $N_e$ (see Eq. \ref{eqrsii}) make challenging the measurement of these quantities. As an example, we show in Fig. \ref{esempi} (top panel) the spectrum of a source for which it is possible to well model the sulfur lines, separating OC and NC, but with faint [O {\small III}]$\lambda$4363. In this case, therefore, it is possible to derive $R_{[S II]}$ but not $R_{[O III]}$. The bottom panel of Fig. \ref{esempi} shows instead how also the vicinity of sulfur and oxygen lines with broad line region (BLR) Balmer emission H$\alpha$ and H$\gamma$ in type 1 AGNs makes the situation more complicated.
These are the reasons for which, for example, \citet{Vaona2012} derived reliable estimates of the electron temperature of the unperturbed NLR gas only for $\sim$ 500 objects, starting from a parent sample of $\sim$ 2500 SDSS AGNs. 

The fact that the OC can be fainter that the unperturbed narrow components (see Paper I, fig. C1), makes further difficult to derive such diagnostic informations for the outflowing ionized material.
Only for a handful of previous studies it was possible to derive, although with large uncertainties, such physical properties. These works are generally based on single luminous targets (e.g. \citealt{Brusa2016,Perna2015a}) or, in the best cases, on small number of sources (e.g. \citealt{Westmoquette2012}). \cite{Harrison2012} used a stacked spectrum of z $\sim 2.4$ ultra-luminous galaxies to estimate the electron density. \cite{Genzel2014}, assuming a pressure equilibrium between outflowing gas and the ionized gas in star-forming regions, proposed for the outflowing gas an electron density of $\sim$ 80 cm$^{-3}$, as derived from star-forming ionized gas in the disks and centres of star-forming galaxies at z $\sim 2$ (see also \citealt{Kaasinen2017}). These few results point to large ranges of values for the electron densities, from $\approx$ 10$^2$ to $>$ 10$^3$ cm$^{-3}$ (see, e.g. \citealt{Rodriguez2013}). 

To the best of our knowledge, the electron temperature has been derived only for six targets (those presented by \citealt{Villar2014,Nesvadba2008} and \citealt{Brusa2016}), with $T_e \approx 1.5\times 10^4$ K.

\begin{figure*}[t]
\centering
\includegraphics[width=17.5cm,angle=180,trim=35 190 15 0,clip]{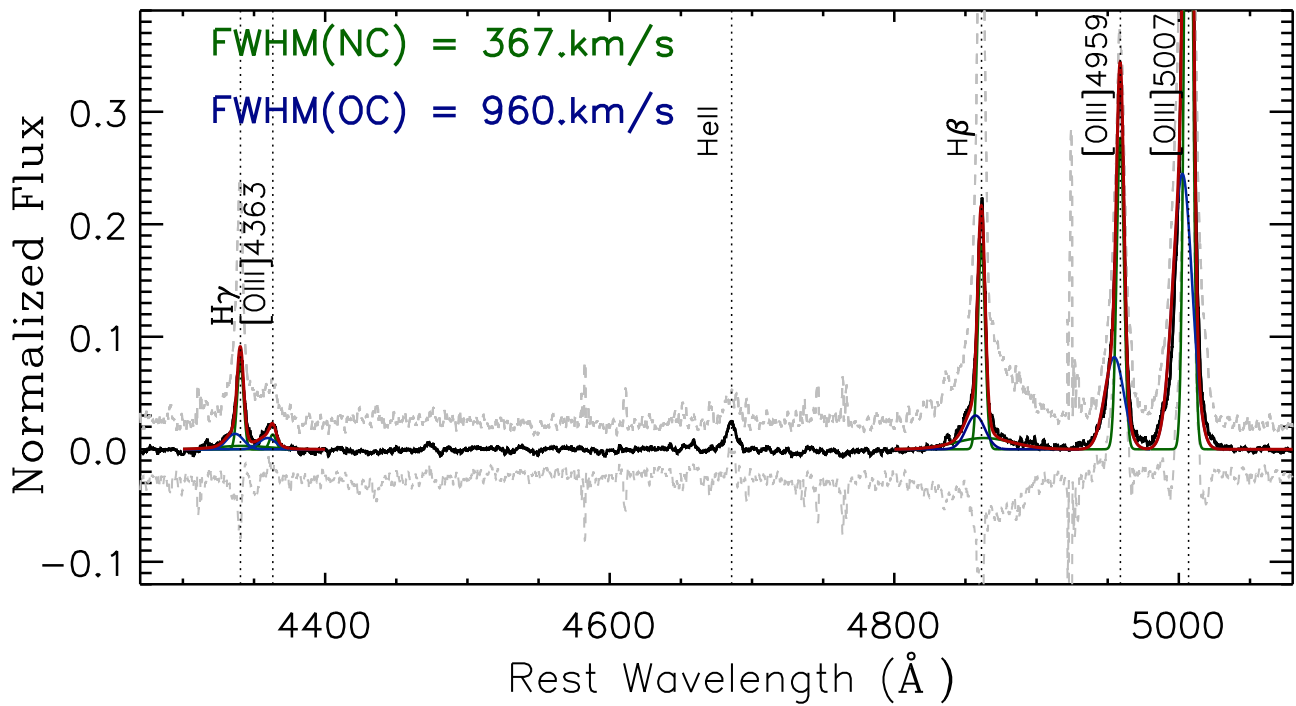}
\caption{\small Median stacked spectrum around the doubly ionized oxygen emission lines obtained combining the spectra of faint/obscured sources with evidence of ionized outflows (see the text for more details). Dashed curves highlight 1$\sigma$ uncertainties. Red curve represents the best-fit result we obtained fitting simultaneously the emission lines displayed. Green Gaussian represent NC and BC components, blue profiles mark OC.}
\label{stackoiii}
\end{figure*}


\begin{figure*}[t]
\centering
\includegraphics[width=18cm,trim=0 150 0 80,clip]{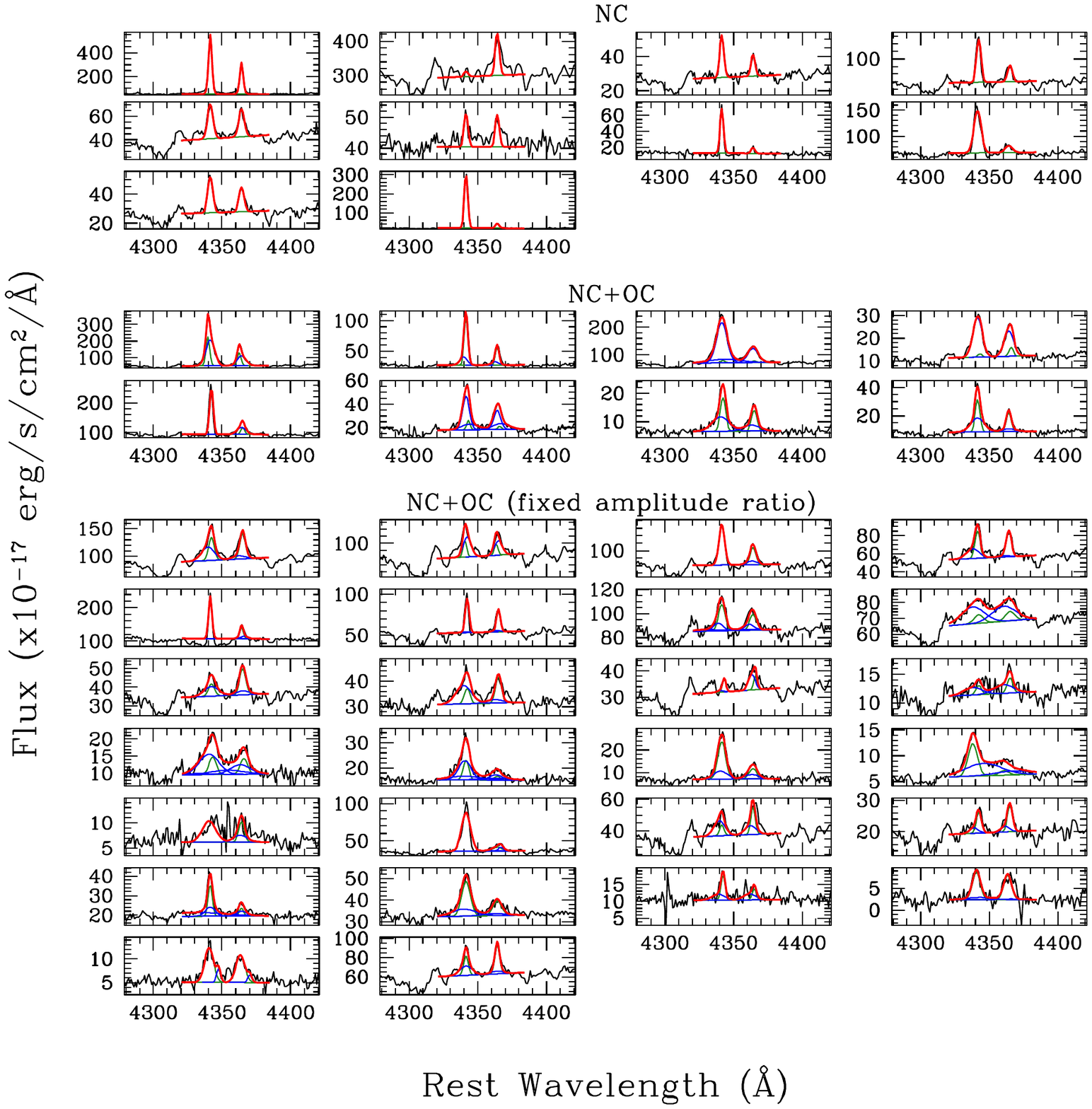}
\caption{\small Zoom in the region of [O {\small III}]$\lambda$4363-H$\gamma\lambda$4342 lines for the 44 sources for which we could determine $R_{[O {\small III}]}$ flux ratios. For each spectrum, the best-fitting components presented in Section \ref{temperaturesection} are superimposed: solid green curves represent the systemic component (NC); blue curves the OC.  }
\label{4363fit}
\end{figure*}

\subsection{Electron Temperature diagnostics}\label{temperaturesection}

In order to constrain at best the electron temperature in ionized gas in our X-ray/SDSS selected AGN sample, we use two different approaches. 
With the first approach, we construct a stacked spectrum to derive temperature-sensitive diagnostic ratio $R_{OIII}$ for both narrow and outflow emitting gas. 
Then, we consider well selected individual objects and derive the same quantity to verify the reliability of the results found with the former approach.

Our X-ray/SDSS sample shows a large variety of AGN types, from typical type 1 AGNs with blue spectra and prominent BLR lines to reddened and/or faint AGNs  (see Paper I, fig. 3). 
Because of the not-so-large number of X-ray/SDSS sources, instead of stacking galaxies in bins of AGN luminosity and/or obscuration, we choose to stack continuum-subtracted spectra.
We consider only those sources for which we were able to model the continuum over the entire wavelength range covering the separation between the doubly ionized oxygen lines  using pPXF best-fit technique (Paper I).
From these, we construct the stacked spectrum combining the continuum-subtracted spectra of all those sources with evidences of outflow in the [O {\small III}] line (75 AGNs). These sources are associated with faint/obscured AGNs ($10^{40} <$ L$_{[O III]} < 10^{42.5}$, with a median luminosity of  $10^{41.2}$ erg/s); this allows us to reduce the possible H$\gamma\lambda$4342 BLR emission in the vicinity of [O {\small III}]$\lambda$4363. 

Figure \ref{stackoiii} shows the median stacked spectrum and the best-fit results we obtain fitting simultaneously all the prominent features in the displayed wavelength range and following the strategy presented in Paper I. The presence of outflow components with FWHM $\approx 1000$ km/s is noticeable in all displayed emission lines. 
From simultaneous multicomponent fit, we derive the flux ratios  $R_{[OIII]}(NC)_{stack}=100_{-30}^{+60}$ and $R_{[OIII]}(OC)_{stack}=33_{-8}^{+74}$ (the errors are computed employing a bootstrap method; \citealt{Peterson2004}).

The stacked spectrum allows us also to check for the possible presence of [Fe {\small II}] contamination of [O {\small III}]$\lambda4363$. In fact, \cite{Curti2017} found that faint iron lines at 4288$\AA$ and 4360$\AA$ concomitantly emerge in high-metallicity stack galaxy spectra, resulting in an overestimation of the oxygen line flux. The absence of [Fe {\small II}]$\lambda$4288 permit to reasonably exclude the possible contribution of iron contamination of oxygen 4363$\AA$ line.

With the second approach, we carefully select a sample of sources with well detected and unblended [O {\small III}]$\lambda$4363 line from the total X-ray/SDSS sample. 
To allow the analysis of the outflow wings in such faint emission line we narrow down the sample selecting only those targets with well detected (S/N$>10$) [O {\small III}]$\lambda$4363. Furthermore, to mitigate blending problems, we discard all those sources with broad H$\gamma$ BLR profiles after a visual inspection. This reduced the sample to 8 sources.
For each target, we fit simultaneously the oxygen line at $4363\AA$ and the Balmer line at $4342\AA$ imposing the same systemics, widths and sets of Gaussian components as obtained from the simultaneous fit in the H$\alpha$ and H$\beta$ regions (Paper I). From such analysis, we derive $R_{OIII}$ flux ratios for both the NC and OC components.

For sources without evidences of outflows in [O {\small III}]$\lambda5007$, the spectral analysis does not require any decomposion between NC and OC emission; we  can therefore relax the requirement on the S/N, imposing a S/N $>5$, and derive temperature-sensitive flux ratios also for this sub-sample.  In this way we obtain $R_{[O III]}(NC)$ ratios for additional 10 AGNs.
The spectra around the region of [O {\small III}]$\lambda$4363 and the fit results for these $8+10$ sources are reported in Fig. \ref{4363fit}.

\begin{figure*}[t]
\centering
\includegraphics[width=8.2cm,trim=0 130 0 100,clip]{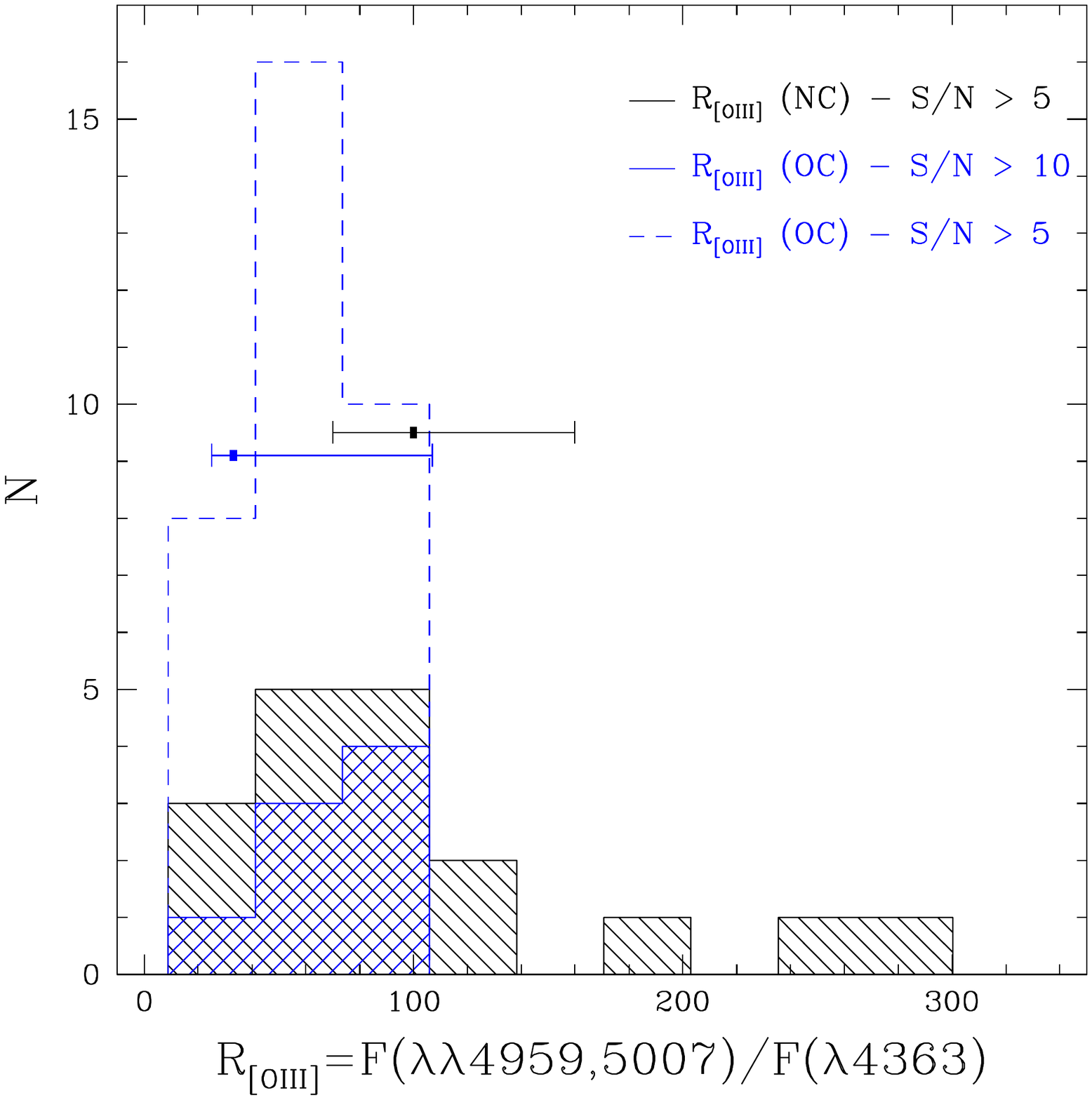}
\hspace{0.01cm}
\includegraphics[width=8.2cm,trim=0 130 0 100,clip]{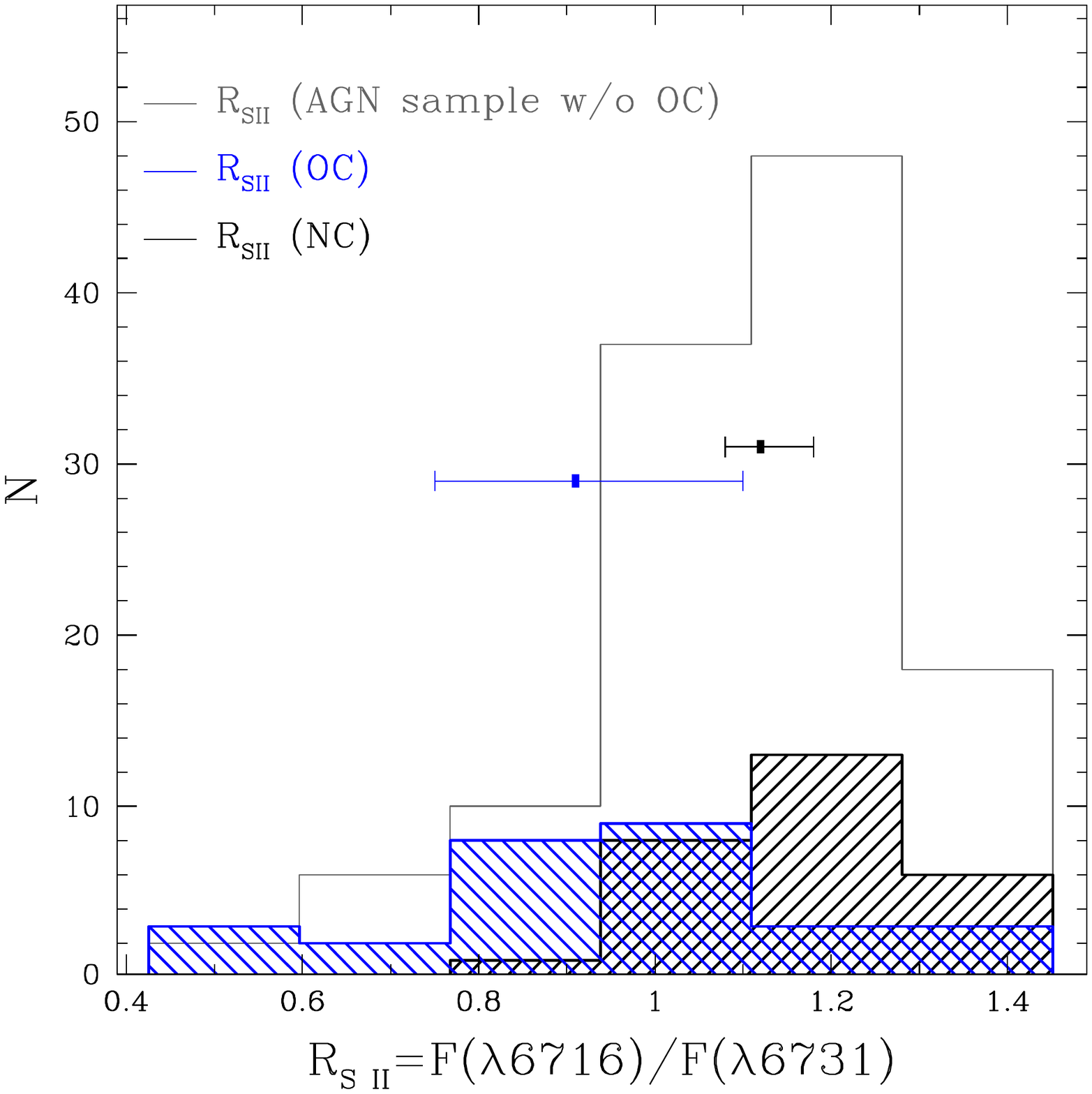}
\caption{\small 
({\it left:}) [O {\small III}] ratio distributions. Grey shaded area marks the $R_{[O {\small III}]}$ distribution for the unperturbed ionised gas (NC). Blue shaded area denotes the outflow emission $R_{[O {\small III}]}$ distribution of S/N$>$10 sources. The blue dashed line show the $R_{[O {\small III}]}$ histogram of the outflow components of all S/N$>$5 sources (see the text for detailed analysis description). In each panel, median and 68\% confidence intervals obtained from the analysis of stacked spectra are also shown for comparison (black and blue symbols, representing line ratios associated with NC and OC, respectively).
({\it right:}) [S II]$\lambda\lambda$6716,6731 ratio distributions. The grey solid line mark the distribution for the AGN sample without evidence of outflows from our line fitting routine. 
The black and blue shaded areas denote the distributions for NC and OC sulfur ratios obtained from the sub-sample of 28 AGNs. 
}
\label{Rdistrib}
\end{figure*}

The derived $R_{OIII}$ distributions for NC (18 AGNs) and OC (8 AGNs) are shown in Fig. \ref{Rdistrib} (left panel; black and blue shaded areas, respectively).

We note that the $R_{OIII}(OC)$ distribution is located closely around the median position of that of $R_{OIII}(NC)$ (i.e., $R_{OIII} \sim$ 90). This could suggest that, on average, NC and OC share similar electron temperatures (Eq. \ref{eqoiii}). These samples are however really small to point toward any conclusion.
We therefore tested this thesis using additional 26 targets with [O {\small III}]$\lambda$4363 detected with 5 $<$ S/N $<10$  and showing evidence of ionized outflows. If we assume the same temperature for both outflowing and systemic ionized gas, the amplitude fractions OC/NC should be the same in [O {\small III}]$\lambda$5007 and [O {\small III}]$\lambda$4363 (see Eq. \ref{eqoiii}). We fitted the emission lines with this additional constrain. 
The fit results are shown in the last part of Fig. \ref{4363fit}. 
All the temperature-sensitive flux ratios are reported in Table \ref{table3}.

We note that the profiles are generally well reproduced under the assumption that $T_e(NC)=T_e(OC)$. Of course, the low quality of the spectra does not allow a strong result significance. 
In Fig. \ref{Rdistrib}, left, we show with dashed blue curve the distribution obtained adding these 26 targets to the outflow sample. In the figure, we also show the results from the stack analysis; the overlap between 68\% confidence intervals for NC and OC flux ratios (black and blue bars, respectively) also tends to support our assumption in deriving the distribution of flux ratio measurements from single targets. 
From the final distribution, we derive the median value $\langle R_{[O III]}(NC) \rangle= \langle R_{[O III]}(OC) \rangle =55\pm28$, with the uncertainty defined by the 68\% confidence interval. 

All median flux ratios obtained so far are collected in Table \ref{table}.

\subsection{Electron density diagnostics}\label{densitysection}

As in the previous section, we use two different approaches to derive the electron density of AGN-ionized gas. 

\begin{figure}[t]
\centering
\includegraphics[width=9.5cm,angle=180,trim=20 0 5 0,clip]{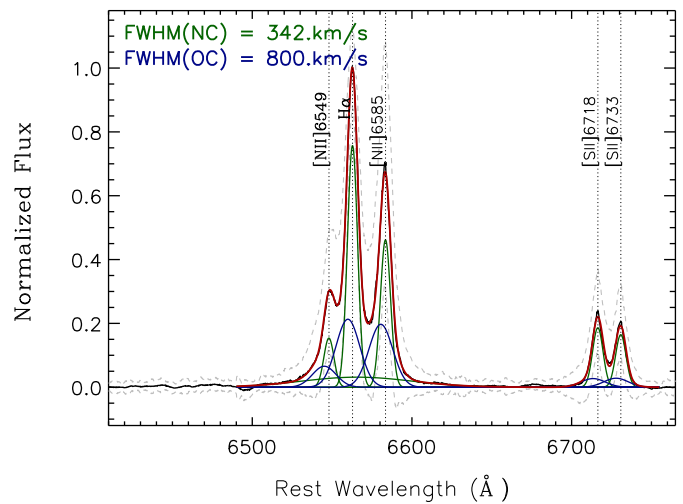}
\caption{\small 
Median stacked spectrum around the [S {\small II}] emission line doublet obtained combining the spectra of sources with evidence of ionized outflows and without blending problems with H$\alpha$ BLR emission.  Dashed curves highlight 1$\sigma$ uncertainties. Red curve represents the best-fit result we obtained fitting simultaneously H$\alpha$, [N II] and [S {\small II}] doublets. Green Gaussian represent NC and BC components, blue profiles mark OC. 
}
\label{stacksii}
\end{figure}

We first construct a stacked spectrum combining the spectra of all those sources with evidence of ionized [O {\small III}]$\lambda5007$ outflows and without blending problems between sulfur and H$\alpha$ BLR emission. For each spectrum satisfying such conditions (90 AGNs), we subtract the continuum emission and normalise the fluxes to the H$\alpha$ peak. Figure \ref{stacksii} shows the median stacked spectrum obtained combining the normalised, continuum-subtracted spectra. 
In order to best model the sulfur profile and distinguish between systemic and perturbed emitting gas, we fit simultaneously H$\alpha$, [N {\small II}] and [S {\small II}] lines (see Paper I for further details). We clearly reveal and separate OC, with FWHM $\approx 800$ km/s, from a narrower unperturbed component, with FWHM of $\approx 350$ km/s, in all emission lines. From the best-fit results (shown in Fig. \ref{stacksii}), we derive  $R_{[SII]}(NC)_{stack}=1.12_{-0.04}^{+0.06}$ and $R_{[SII]}(OC)_{stack}=0.91_{-0.19}^{+0.16}$ (as before, the errors are computed employing a bootstrap method).

The second approach employs the analysis of optical spectra of single objects. 
As a first step, we measure the NLR diagnostic ratio $R_{[SII]}$ for all AGNs with no severe and ambiguous blending with H$\alpha$ BLR emission and without signatures of outflows revealed in simultaneous fits, resulting in a sample of 121 AGNs. In fact, when OC components are revealed, the doublet lines are usually severely blended and, in general, the fitting procedure give ambiguous results (see, e.g. \citealt{Perna2015a,Rodriguez2013,Villar2014}).
 From this sample we obtained the distribution shown in Fig. \ref{Rdistrib}, right (grey histogram).

To study the electron density of the outflowing regions, we focus the analysis on those AGNs with the simplest spectral profiles, i.e. with well defined [S {\small II}] wings modelled with two kinematic components (NC $+$ OC) and, as before, without strong blending with BLR emission. In Paper I we noticed that about 40\% of X-ray/SDSS AGNs of our parent sample display signatures of ionized outflows; the above mentioned conditions are however satisfied by only 28 sources. This reflects the difficulties in the electron density measurements. The fitted spectra are shown in Fig. \ref{fit}. From this sample, we compute $R_{[SII]}$ ratios for both NC and OC. The $R_{[SII]}(NC)$ distribution (Fig. \ref{Rdistrib}, right; black shaded area) has a smaller spread when compared with that of the AGN sample without OC components, because of the particular selection. Instead, the $R_{[SII]}(OC)$ distribution (blue area) covers a larger range of values and is peaked at lower ratios. From these distributions, we derive the median values $\langle R_{[SII]}(NC) \rangle=1.16_{-0.17}^{+0.16}$ and $\langle R_{[SII]}(OC) \rangle =0.94_{-0.24}^{+0.27}$ (the errors define 68\% confidence intervals; see Table \ref{table2} for the compilation of all flux ratio measurements).


From each [S {\small II}] intensity ratio, in principle, we could derive an estimate of the electron density from Eq. \ref{eqrsii}, taking into account the dependence on the electron temperature. This means that for each source, in order to derive at best $N_e$, we should be able to detect and analyse all the emission lines needed for electron density {\it and} temperature diagnostics. Unfortunately, the faintness of the temperature-sensitive emission line [O {\small III}]$\lambda$4363 does not allow spectral analysis for all but 7 of our targets selected to study the [S {\small II}] emission (see, e.g. the spectrum in Fig. \ref{esempi}, panel a). Therefore, we choose to follow a statistical approach, and derive median electron densities of outflowing and systemic gas for given median electron temperatures. 

\begin{table*}
\centering
\caption{Plasma properties}
\begin{minipage}[t]{12cm}
\footnotesize
\centering
\begin{tabular}{|c|ccc|ccc|}
\cline{2-7}
\multicolumn{1}{ c |}{} & \multicolumn{3}{c|}{stacked} & \multicolumn{3}{c|}{individual}\\ 
\multicolumn{1}{ c |}{} & NC & OC & ($\#$) & NC & OC & ($\#$)\\
\multicolumn{1}{ c |}{} & {\scriptsize (1)} & {\scriptsize (2)}  &{\scriptsize (3) }     & {\scriptsize (4)}  &{\scriptsize (5)} &{\scriptsize (6)}\\
\hline
R$_{[S II]}$ & 1.12$_{-0.04}^{+0.06}$ & 0.91$_{-0.19}^{+0.16}$ & (90) & $1.16_{-0.17}^{+0.16}$ & $0.94_{-0.24}^{+0.27}$ & (28) \\
R$_{[O III]}$ &$100_{-30}^{+60}$  &$33_{-8}^{+74}$ &(75) & $55_{-28}^{+28}$ & $55_{-28}^{+28}$ &(34) \\
\hline \hline
$N_e$ & $550_{-130}^{+110}$ cm$^{-3}$&$1200_{-500}^{+1500}$ cm$^{-3}$& (90) &$480_{-300}^{+450}$ cm$^{-3}$ & $1100_{-750}^{+1900}$ cm$^{-3}$ & (28)\\
$T_e$ & $1.3_{-0.2}^{+0.2} \times 10^4$ K & $2.3_{-1.0}^{+0.6}\times 10^4$ K&(75) & $1.7_{-0.3}^{+1.1}\times 10^4$ & $1.7_{-0.3}^{+1.1}\times 10^4$ K & (34)\\
\toprule
\end{tabular}
\label{table}
\end{minipage}
\tablefoot{Median estimates for sulfur and oxygen line ratios (first two rows) and electron density and temperature (third and fourth rows), for both narrow (NC) and outflow components (OC). Columns (1) and (2) show the results obtained from the analysis of median stacked spectra constructed collecting \# [column (3)] spectra. Columns (4) and (5) display the median results obtained analysing sub-samples of \# [column (6)] individual objects. The oxygen flux ratio and electron temperature obtained from the analysis of individual targets are obtained under the assumptions described in the text, and are referred to both NC and OC.}
\end{table*}

\subsection{Results: the plasma properties}\label{resultsection}

We compute fiducial estimates of the electron temperatures of both NC and OC through Eq. \ref{eqoiii}, using the median doubly ionized oxygen flux ratios derived from stack analysis, 
$T_e(NC)=(1.3 \pm 0.2)\times 10^4$ K and $T_e(OC)=2.3_{-1.0}^{+0.6}\times 10^4$ K. \\
We also use the median value of the final $R_{[O {\small III}]}$ distribution (Fig. \ref{Rdistrib}, left, blue dashed histogram), to derive an additional estimate of the electron temperature in the outflow region, $T_e(OC)=1.7_{-0.3}^{+1.1}\times 10^4$ K. Because of the similarities in the distributions of NC and OC, and the fact that this temperature value is also consistent within $\sim1\sigma$ with the stack analysis measurements, we consider  $T_e=1.7\times 10^4$ K  as a median electron temperature for the entire NLR emitting gas (i.e., NC and OC emitting material). 

We therefore adopt such median temperature to derive fiducial electron densities from Eq. \ref{eqrsii}, for both best-fit NC and OC results obtained from stack analysis: 
\begin{equation*}
N_e^{stack} (NC)=550_{-110}^{+130}\ cm^{-3}, 
\end{equation*} 
\begin{equation*}
N_e^{stack} (OC)=1200_{-500}^{+1500}\ cm^{-3}. 
\end{equation*} 

In the same manner, we also derive fully consistent electron densities from the median values of sulfur ratio distributions: $\langle N_e (NC) \rangle =480_{-300}^{+450}$ cm$^{-3}$ and $\langle N_e (OC) \rangle=1100_{-750}^{+1900}$ cm$^{-3}$.\footnote{Similar results are obtained for the seven sources for which we were able to analyse both doubly ionized oxygen lines and sulfur lines (see Sect. \ref{densitysection}): $\langle N_e (NC) \rangle =510 $ cm$^{-3}$ and $\langle N_e (OC) \rangle = 1100$ cm$^{-3}$.}

The targets we used to construct the stacked spectra have observed [O {\small III}]$\lambda$5007 luminosity range of $10^{40} \lesssim  L_{[O III]} \lesssim 10^{42.5}$, with a median luminosity of  $10^{41.2}$ erg/s, similarly to those of the individual spectra reported in Table \ref{table2} and \ref{table3}. Because of their faint/obscured nature, we also report the median values of doubly ionized luminosity corrected for the extinction,  $L_{[O III]}^{int}\sim 10^{41.6}$ erg/s (derived using Balmer decrement arguments and assuming Case B ratio of 3.1 and the SMC dust-reddening law; \citealt{Perna2015a}) and of the intrinsic X-ray luminosity, $L_{X}\sim 10^{42.5}$ erg/s. The limitation in AGN luminosity regime is due to the above mentioned conditions required to analyse [S {\small II}] and [O {\small III}]$\lambda$4363 which, for instance, preclude the inclusion of the majority of unobscured type 1 AGNs. Therefore, we note that the results presented here may be relevant only to characterise the outflow plasma properties of faint and moderately luminous AGNs ($10^{40.5}\lesssim L_X\lesssim 10^{44}$, or $10^{41.2}\lesssim L_{[O III]}^{int} \lesssim 10^{42}$). 

Narrow component plasma properties have been derived for larger samples of SDSS Seyferts by other authors in the past decade (e.g. \citealt{Zhang2013,Vaona2012}). Our estimates are totally consistent with the median values indicated by these authors ($N_e \approx 400$ and $T_e \approx 1.5\times10^4$ K).

Although with large uncertainties, the outflow plasma condition estimates presented in this work are, to the best of our knowledge, the first average estimate from stack analysis and medium-size samples ($\approx$ 30 targets) of AGNs. 
Moreover, despite the fact that our median values for $N_e(NC)$ and $N_e(OC)$ are still comparable within the errors,  
our analysis suggests that the outflowing gas may actually be characterised by a large range of electron densities, possibly favouring the high-density regime.
As a consequence, the most common assumption to derive outflow energetics ($N_e=100$ cm$^{-3}$) may overestimate such measurements even by a factor of 10.

\begin{figure*}[t]
\centering
\includegraphics[width=15cm,trim=0 130 0 80,clip]{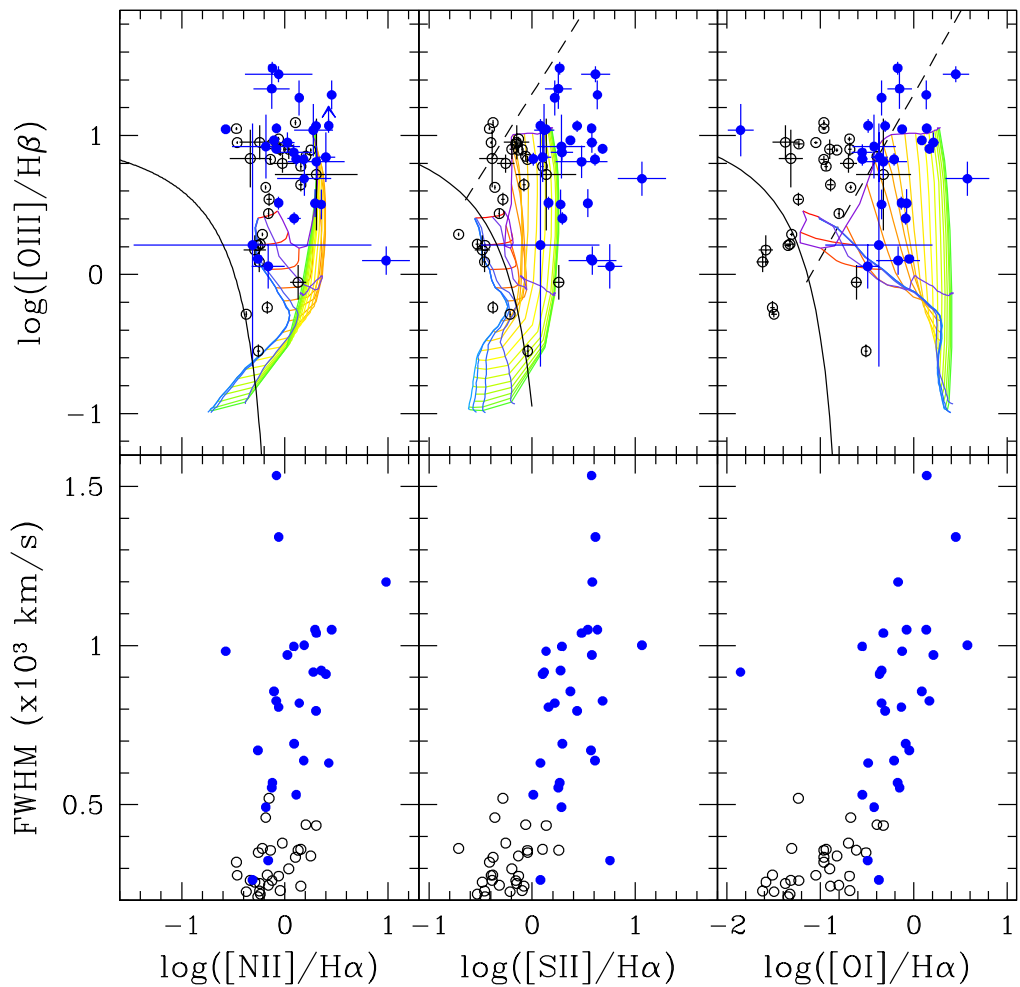}
\caption{\small ({\it Top panels}:) [O {\small III}]/H$\beta$ versus [N {\small II}]/H$\alpha$, [S {\small II}]/H$\alpha$ and [O I]/H$\alpha$  BPT diagrams for the 28 sources used to derive the outflow electron density. Solid and dashed black lines represent the curve used to separate the purely SF galaxies, AGNs and LINER loci. Shock model grids are overplotted, with increasing velocities, from 100 to 1000 km/s (red to green lines), and magnetic field (blue to purple curves). ({\it Bottom panels}:) FWHM plotted against log([N {\small II}]/H$\alpha$), log([S {\small II}]/H$\alpha$), and  [O I]/H$\alpha$ from left to right. Both NC (open black) and OC (solid blue symbols) are shown. Positive correlations are found between the three ionization state and the gas kinematics tracers. 
}
\label{BPT28}
\end{figure*}

\subsection{Possible bias in T$_e$ due to [O {\small III}]$\lambda$4363 selection}\label{bias}

The fact that we are computing the electron temperature for those sources with intense [O {\small III}]$\lambda$4363 could bias the results obtained with the second approach, by favouring targets with higher $T_e$ (Eq. \ref{eqoiii}). 
We compute 1$\sigma$ upper limits for all those sources without clear detection and, from the median value of their $R_{OIII}$ distribution, we derive an upper limit on the electron temperature of $\approx3\times10^4$ K. Unfortunately, this value is not useful to our purpose. However, the fact that we have not observed any difference in the median $T_e$ between the S/N $>10$ and $5<$ S/N $<10$ samples could suggest that the bias is, if any, negligible. The same behaviour can be seen for the large sample studied by \citet{Zhang2013}, for which the same NLR unperturbed gas electron temperature has been found for both their two [O {\small III}]$\lambda$4363 luminosity subclasses of Seyferts. 
Indeed, the scarce relevance of bias may be also stressed by the consistent results we obtained from stack analysis, for which only faint/obscured AGNs have been taken into account.

\section{Outflowing gas ionization mechanisms}\label{mechanisms}

The unambiguous separation between NC and OC components in the [S {\small II}] emission lines for the medium size sample of the 28 AGNs from which we derived electron density estimates, enables the study of the diagnostic diagrams [O {\small III}]/H$\beta$ versus [S {\small II}]/H$\alpha$ (see, e.g.  \citealt{Kewley2006}). Furthermore, we find well detected [O {\small I}]$\lambda$6300 emission lines in all sources (see Fig. \ref{fit}). Therefore, taking advantage from simultaneous analysis results, we fit this faint emission constraining the systemics and the FWHM of NC and OC components and derive the intensity ratio needed for a second diagnostic diagram,  [O {\small III}]/H$\beta$ vs. [O {\small I}]/H$\alpha$ (\citealt{Kewley2006}).

Figure \ref{BPT28} shows three diagnostic diagrams, [O {\small III}]/H$\beta$ versus [N {\small II}]/H$\alpha$ (already presented in Paper I for the entire sample of X-ray/SDSS AGNs), [S {\small II}]/H$\alpha$ and [O  {\small I}]/H$\alpha$ for both NC (open black symbols) and OC (blue circles). The lines drawn in the diagrams correspond to the optical classification scheme of \citet{Kewley2006,Kewley2013}: in the second and third diagram, the diagonal line marks the LINER locus.  
The OC appear associated with the same level of ionization of NC (i.e., same [O {\small III}]/H$\beta$ ratios) and similar [N {\small II}]/H$\alpha$ but larger [S  {\small II}]/H$\alpha$ and [O  {\small I}]/H$\alpha$ ratios. The second and third diagnostic diagrams clearly point to a LINER-like emission for the outflow components. Such kind of line ratios are generally associated with ionization by fast radiative shocks (e.g. \citealt{Allen2008}; but see also \citealt{Belfiore2016}). Shock model results have been made available for a large range of physical parameters: pre-shock density N$_e^{pre}$ (to be distinguished by electron density we measured previously, and possibly associated with 'post-shock' regions; see, e.g. \citealt{Harrison2012}), shock velocity, magnetic field and abundances. 
We superimposed on the figure a grid of shock model with assumed solar abundance and a pre-shock density of 100 cm$^{-3}$ (ITERA; \citealt{GrovesAllen2010}). The grid shows different line ratios for various values of magnetic field and shock velocities (up to 1000 km/s).
The models, however, fail to reproduce the exact position of the OC line ratios in the BPT diagrams. We tested all available shock models from the ITERA library without any improvement. 

It is possible that more extreme set of parameters are needed to reproduce the shocks in AGN-driven outflows: under particular assumptions, shock velocities and  gas velocity dispersion can be similar, and therefore even larger than 1000 km/s (see \citealt{McElroy2015} discussion). Radiation pressure-dominated photo-ionization models easily reproduce the loci occupied by NC measurements (see, the model grids in \citealt{Westmoquette2012}) but, as the shock models, fail to cover the highest [O {\small I}]/H$\alpha$ and [S  {\small II}]/H$\alpha$ ratios. We notice however that these values might also be related to the presence of high metallicity regions within outflowing gas (see \citealt{Villar2014}), and that further observational efforts and theoretical investigation are required to discriminate between radiation pressur and shock models.

Another strong evidence for shock excitation interpretation is given by the observed correlation between the gas kinematics and ionization state (\citealt{DopitaSutherland1995}). The line ratios produced in photo-ionized regions should be independent of the gas kinematics, while are expected to correlate with the kinematics of the shock-ionized material (see \citealt{McElroy2015,Ho2014,Arribas2014}).
In the lower panels of Fig. \ref{BPT28} we show the FWHM against [N {\small II}]/H$\alpha$, [S  {\small II}]/H$\alpha$ and [O  {\small I}]/H$\alpha$ ratios for both NC and OC. 
We note that the FWHM should not suffer for strong degeneracy between the [S {\small II}] doublet and [N {\small II}]-H$\alpha$ complex components, because derived from our simultaneous fits (Fig. \ref{fit}). We use Spearman rank correlation coefficients to determine the significance of the observed trends between line ratios and velocities. We find, on average, coefficients of $\approx$ 0.45 with probabilities of $\lesssim$ 0.01 for the correlation being observed by chance, for both NC and OC separately. The same correlation  has been found by \citet{Arribas2014} for narrow components (but not for the outflow components), confirming the complex kinematic conditions within the NLR.

\section{Discussion and Conclusions}\label{sdssdiscussion}

In \citet{Perna2017}, we analysed SDSS optical and X-ray spectra of a sample of 563 AGN at  z $<0.8$. 
In this paper, we have taken advantage of the optical spectra modelling presented in Paper I to  derive the plasma properties for both narrow line and outflowing component emission. For the first time we derived estimates of electron temperature and density of the outflowing gas from stack analysis and medium size samples ($\approx 30$) of faint and moderately luminous AGNs. 

 Studying individual AGN spectra, we found indications suggesting that $T_e$ may be quite similar in both outflowing and unperturbed gas, with a median value $T_e=1.7_{-0.3}^{+1.1}\times 10^4$ K. The analysis of a stacked spectrum derived combining the spectra of faint/obscured X-ray detected AGNs allowed the determination of independent estimates for both NC and OC: $T_e(NC)=1.3\times 10^4$ K and $T_e(OC)\approx2.4\times 10^4$ K. These values are still consistent (within 1$\sigma$) with the median value obtained from individual spectra analysis. 

In Sect. \ref{mechanisms} we presented BPT and ionization state - velocity diagnostics diagrams for 28 AGNs selected from the X-ray/SDSS sample; such diagnostics indicate a shock excitation interpretation for the observed ionized outflows. If it is the case, the similar electron temperature of OC and NC could be consistent with theoretical arguments, which postulate that the (forward) shock accelerating the ISM gas is strongly cooled, so that the gas temperature rapidly returns to its pre-shock value (\citealt{King2014}; see also its fig. 2). However, if the bulk of emission is associated to shock excitation, we should observe much higher temperatures ($T_e>5 \times 10^5$ K; \citealt{Osterbrock2006}). These temperatures are ruled out by our observations: in that case, we should observe much stronger [O {\small III}]$\lambda$4363 lines. Even considering possible dust extinction effects, which could affect the doubly ionized oxygen line ratios of a factor of 0.85 (assuming a SMC dust-reddening law; \citealt{Prevot1984}), our temperature estimates would result just $\sim 10\% $ higher.

The position of OC component flux ratios in the BPT may also be associated with different physical origins. For example, the high [S {\small II}]/H$\alpha$ and [O {\small I}]/H$\alpha$ OC ratios  can be associated with hard ionization radiation field, assuming the material is close to the AGN, or with high metallicities (e.g. \citealt{Belfiore2016,Villar2014}). 
Moreover, the kinematic separation between different emission line components does not ensure a common spatial distribution for all the emitting species in diagnostic diagrams. Hence, different emitting species could be characterised by distinct physical conditions. 
Spatially resolved information is required to better investigate the ionization source of outflowing gas.

The outflowing gas electron densities we derived analysing individual targets and a stacked spectrum displayed a wide range of values, with a distribution characterised by a median value of $\approx1200$ cm$^{-3}$ and a 68\% interval going from 700 to 3000 cm$^{-3}$ (Table \ref{table}). 

\citet{Xu2007} studied the NLR emission of $\sim$ 100 SDSS type 1 AGNs. They observed a negative trend between the electron density and the blueshift of the [O {\small III}] wing. This result conflicts with our measurements. However, we note that all these results are currently limited to small number of sources: the trend suggested by \citet{Xu2007} is based on the measurement of low $N_e$ in 6 out 54 targets with signature of outflows.
Moreover, they derived electron density estimates without any separation between narrow and outflow components, making the comparison difficult.

On the other hand, other indications in literature suggest even higher $N_e$ in the outflowing regions: \citet{Villar2015}, studying high-ionization lines such as [Fe X] and [Ne V], proposed electron densities up to $10^5$ cm$^{-3}$. However, such emission lines, although actually tracing outflowing gas (see, e.g. \citealt{Lanzuisi2015}), have high ionization potential and critical densities (IP $\gtrsim100$ eV and $N_c\gtrsim10^7$ cm$^{-3}$, to be compared to sulfur IP $=10.36$ and $N_c\approx 2500$ cm$^{-3}$) and could be associated with more internal regions (see \citealt{Rose2015a,Rose2015b}, in which such lines have been proposed to occupy a region between the BLR and the inner walls of the torus; see also \citealt{Morse1998}). When we consider kpc-scale outflows, typical electron density may be instead more similar to those of narrow emission lines in NLR rather than in the nuclear regions. The slightly higher density ($\approx 1200$ instead of $\approx 500$ cm$^{-3}$) may be explained by a possible compression due to the AGN wind on the ISM material. 

Summarising, all these considerations, based on the results we obtained and on speculative arguments, suggest a more conservative approach in the estimate of the outflow energetics:  the most typical assumption in deriving crucial outflow energetics (i.e., $N_e=100$  cm$^{-3}$) may, in fact, overestimate the outflow energetics of a factor up to 10.

\vspace{2cm}

{\small 
{\it Acknowledgments:} MP, GL and MB acknowledge support from the FP7 Career Integration Grant ``eEASy'' (``SMBH evolution through cosmic time: from current surveys to eROSITA-Euclid AGN Synergies'', CIG 321913). GL acknowledges financial support from ASI-INAF I/037/12/0.
Support for this publication was provided by the Italian National Institute for Astrophysics (INAF) through 
PRIN-INAF-2014  (``Windy  Black  Holes
combing  galaxy  evolution''). We thanks the anonymous referee for his/her constructive comments to the paper.
MP thanks A. Citro and S. Quai for useful discussion on stacked spectra analysis.
Funding for the Sloan Digital Sky Survey (SDSS) has been provided by the Alfred P. Sloan Foundation, the Participating Institutions, the National Aeronautics and Space Administration, the National Science Foundation, the U.S. Department of Energy, the Japanese Monbukagakusho, and the Max Planck Society. The SDSS Web site is \url{http://www.sdss.org/}.

    The SDSS is managed by the Astrophysical Research Consortium (ARC) for the Participating Institutions. The Participating Institutions are The University of Chicago, Fermilab, the Institute for Advanced Study, the Japan Participation Group, The Johns Hopkins University, Los Alamos National Laboratory, the Max-Planck-Institute for Astronomy (MPIA), the Max-Planck-Institute for Astrophysics (MPA), New Mexico State University, University of Pittsburgh, Princeton University, the United States Naval Observatory, and the University of Washington.
}

\begin{appendix}

\section{X-ray/SDSS sub-samples}\label{AppendixA}

In Fig. \ref{fit} we report the multicomponent simultaneous fit results of the 28 sources for which it was possible to well disentangle between different kinematic components in the forbidden doublet of singly ionized sulfur (Sect. \ref{densitysection}). The flux ratios R$_{[S II]}$ derived for the entire set of 28 AGNs are reported in Table \ref{table2}; the doubly ionized oxygen ratios obtained for the sub-samples of AGNs defined in Sect. \ref{temperaturesection} and used to construct the histograms in Fig. \ref{Rdistrib} are reported in Table \ref{table3}.

\begin{table*}[h]
\centering
\caption{Plasma diagnostics sub-samples}
\label{table2}
\begin{tabular}{ccccccc}
SDSS name & ID-MJD-FIBER & z &sample & L$_{[O III]}$ & \multicolumn{2}{c}{R$_{[S II]}$}\\
 {\scriptsize (1)} & {\scriptsize (2)}  &{\scriptsize (3) }     & {\scriptsize (4)}  &{\scriptsize (5)} &{\scriptsize (6)}&{\scriptsize (7)}\\
\toprule
J135317.8+332927  & 2117-54115-0351 & 0.0079 & T13 &  39.88 &1.06$\pm$0.05&0.92$\pm$0.14\\
J125725.2+272416    & 2241-54156-0195 & 0.0161 & G11 & 39.49 &1.23$\pm$0.40&1.16$\pm$0.28\\
J153457.2+233013       & 2163-53823-0058 & 0.0184 & G11 & 39.19 &1.38$\pm$0.55&1.33$\pm$0.15\\
J121049.6+392822  & 1995-53415-0205 & 0.0226 & G11 & 39.94  &0.96$\pm$0.09&0.95$\pm$0.15\\
J120429.6+201858       & 2608-54474-0555 & 0.0226 & G11 & 41.17  &0.99$\pm$0.05&0.67$\pm$0.10\\
J130125.2+291849  & 2011-53499-0492 & 0.0234 & T13 &   40.66 &1.16$\pm$0.40&0.93$\pm$0.45\\
J104451.7+063548    & 1000-52643-0080 & 0.0276 & G11 & 40.98  &1.00$\pm$0.04&0.89$\pm$0.09\\
J123651.1+453904  & 1372-53062-0412 & 0.0303 & T13 & 40.67  &1.34$\pm$0.07&0.81$\pm$0.08\\
J160515.8+174227  & 2200-53875-0423 & 0.0316 & G11 & 40.41  &1.13$\pm$0.21&0.95$\pm$0.31\\
J144921.5+631614    & 0609-52339-0531 & 0.0417 & G11 & 41.17 &0.97$\pm$0.11&1.25$\pm$0.10\\
J155855.7+024833    & 0595-52023-0179 & 0.0468 & G11 & 40.85  &1.26$\pm$0.08&0.45$\pm$0.08\\
J135602.6+182217  & 2756-54508-0198 & 0.0506 & T13 & 41.11  &1.24$\pm$0.50&0.77$\pm$0.50\\
J002920.3-001028   & 0391-51782-0155 & 0.0605 & G11 & 40.84  &0.98$\pm$0.13&0.79$\pm$0.08\\
J125558.7+291459  & 2011-53499-0244 & 0.0681 & T13 &  41.13 &1.12$\pm$0.60&1.07$\pm$0.60\\
J125830.1-015837  & 0338-51694-0480 & 0.0803 & T13 &  41.18 &1.14$\pm$0.06&1.03$\pm$0.10\\
J081444.2+363640  & 0892-52378-0579 & 0.0825 & T13 & 40.80  &1.33$\pm$0.13&0.81$\pm$0.22\\
J151141.2+051809  & 1832-54259-0119 & 0.0842 & T13 & 41.71  &1.12$\pm$0.35&0.59$\pm$0.18\\
J124415.2+165350   & 2601-54144-0282 & 0.0876 & G11 & 40.84  &1.34$\pm$0.23&0.94$\pm$0.18\\
J143602.5+330754 & 1841-53491-0584 & 0.0938 & T13 &  41.36 & 1.15$\pm$0.05&1.10$\pm$0.36 \\
J000703.6+155423  & 0751-52251-0577 & 0.1141 & T13 & 41.94  &1.16$\pm$0.55&1.22$\pm$0.60\\
J111847.0+075419   & 1617-53112-0469 & 0.1269 & G11 & 41.48  &0.77$\pm$0.26&0.47$\pm$0.11\\
J090935.5+105210  & 1739-53050-0421 & 0.1654 & T13 &  42.33 &1.04$\pm$0.23&1.06$\pm$0.20\\
J141602.1+360923   & 1643-53143-0153 & 0.1708 & G11 & 41.50  &1.32$\pm$0.17&1.29$\pm$0.25\\
J085331.0+175339    & 2281-53711-0179 & 0.1865 & G11 & 42.56  &1.11$\pm$0.07&1.08$\pm$0.12\\
J121249.8+065945  & 1624-53386-0056 & 0.2095 & W12 &  40.89  &1.16$\pm$0.23&0.83$\pm$0.15\\
J150407.5$-$024816   & 0922-52426-0127 & 0.2169 & G11 & 42.02  &1.24$\pm$0.08&0.97$\pm$0.05\\
J090036.8+205340    & 2282-53683-0103 & 0.2357 & G11 & 42.57  &1.33$\pm$0.10&0.71$\pm$0.13\\
J142314.2+505537  & 1045-52725-0072 & 0.2754 & W12 &  42.89  &1.13$\pm$0.30&1.31$\pm$0.26\\

\toprule
\end{tabular}
\tablefoot{Column (1) and (2): names and MJD, plate and fibre numbers which identify the SDSS targets. Column (3): redshifts. Column (4): parent sample from which X-ray/SDSS targets have been selected (abbreviations: G11, \citealt{Georgakakis2011}; T13, \citealt{Trichas2013}; W12, \citealt{Wu2012}). Column (5): [O III]$\lambda5007$ total luminosity. Columns (6) and (7): NC and OC flux ratios between [S II]$\lambda6716$ and $\lambda6731$ lines. }
\end{table*}

\begin{table*}[h]
\centering
\caption{Plasma diagnostics sub-samples - [O {\small III}] flux ratios}
\label{table3}
\begin{tabular}{ccccccc}
SDSS name & ID-MJD-FIBER & z &sample & L$_{[O III]}$ & \multicolumn{2}{c}{R$_{[O III]}$}\\
 {\scriptsize (1)} & {\scriptsize (2)}  &{\scriptsize (3) }     & {\scriptsize (4)}  &{\scriptsize (5)} &{\scriptsize (6)}&{\scriptsize (7)}\\
\toprule
J122546.7+123942 & 1615-53166-0388 & 0.0086 & G11 & 40.95 &127$\pm$35 & 95$\pm$23\\
J140040.5-015518 & 0915-52443-0437 & 0.0250 & G11 & 41.07 &102$\pm$9 & 19$\pm$2\\
J144012.7+024743 & 0536-52024-0575 & 0.0299 & G11 & 41.16 &71$\pm$3 & 94$\pm$15\\
J103408.5+600152 & 0560-52296-0520 & 0.0510 & G11 & 42.44 &284$\pm$150 & 79$\pm$5\\
J121839.4+470627 & 1451-53117-0190 & 0.0939 & G11 & 42.16 &245$\pm$140 & 61$\pm$4\\
J093952.7+355358 & 1594-52992-0417 & 0.1366 & G11 & 42.34 &95$\pm$15 & 49$\pm$2\\
$^*$J090935.5+105210  & 1739-53050-0421 & 0.1654 & T13 &  42.33 & 50$\pm$4&44$\pm$11\\
$^*$J085331.0+175339  & 2281-53711-0179 & 0.1865 & G11 & 42.56 & 63$\pm$2 &90$\pm$11\\
J122548.8+333248  & 2015-53819-0251 & 0.0011 & G11 & 38.92 &52$\pm$1 & --\\
J134208.3+353915  & 2101-53858-0049 & 0.0036 & G11 & 39.38 &25$\pm$1 & --\\
J112613.7+564809 & 1309-52762-0263 & 0.0101 & T13 & 39.44 & 76$\pm$4 & --\\
J081937.9+210651  & 1927-53321-0261 & 0.0185 & G11 & 40.39 &93$\pm$6 & --\\
J005329.9$-$084604  & 0657-52177-0458 & 0.0190 & G11 & 40.79 &115$\pm$7 & --\\
J104341.2+591653 & 0561-52295-0252 & 0.0285 & T13& 39.68 & 97$\pm$4 & --\\
J235951.7$-$092632 & 0650-52143-0461 & 0.0423 & T13 & 41.09 & 54$\pm$3 & --\\
J080535.0+240950  & 1265-52705-0158 & 0.0597 & G11 & 41.43 &36$\pm$2 & --\\
J134427.5+560128 & 1321-52764-0624 & 0.0706 & T13 & 41.12 & 187$\pm$4 & --\\
J122137.9+043026 & 2880-54509-0318 & 0.0947 & G11 & 40.69 &30$\pm$2 & --\\
\midrule 
$^*$J120429.6+201858  & 2608-54474-0555 & 0.0226 & G11 & 41.17 & 41$\pm$3&41$\pm$3\\
$^*$J130125.2+291849  & 2011-53499-0492 & 0.0234 & T13 &   40.67 & 61$\pm$26&61$\pm$26\\
J112301.3+470308 & 1441-53083-0560 & 0.0252 & G11 & 40.95&  44$\pm$3&44$\pm$3\\
J082443.2+295923  & 1207-52672-0500 & 0.0254 & G11 &41.19&  89$\pm$5&89$\pm$5\\
J080359.2+234520 & 1265-52705-0271 & 0.0294 & G11 &41.08&  58$\pm$4&58$\pm$4\\
J104930.9+225752 & 2481-54068-0247 & 0.0328 & G11 & 41.25 & 59$\pm$3&59$\pm$3\\
J110929.3+284129 & 2213-53792-0223 & 0.0329 & T13 &40.87 & 32$\pm$2&32$\pm$3\\
J095914.7+125916  & 1744-53055-0385 & 0.0343 & G11 &41.47&  86$\pm$10&86$\pm$10\\
J115704.8+524903  & 0882-52370-0430 & 0.0356 & G11 &41.26& 84$\pm$4&84$\pm$4\\
J113549.0+565708 & 1311-52765-0262 & 0.0514 & G11 &41.69& 75$\pm$4&75$\pm$4\\
J151640.2+001501 & 0312-51689-0471 & 0.0526 & G11 & 41.04& 9$\pm$1&9$\pm$1\\
J150754.3+010816 & 0540-51996-0253 & 0.0610 & G11 &41.53& 51$\pm$3&51$\pm$3\\
J151106.4+054122   & 1833-54561-0404 & 0.0806 & G11 &41.61&  49$\pm$2&49$\pm$2\\
J102147.8+131228& 1746-53062-0513 & 0.0852 & G11 & 41.09& 18$\pm$2&18$\pm$2\\
J103014.4+431520 & 1429-52990-0131 & 0.0985 & T13 & 41.65 & 48$\pm$4&48$\pm$4\\
$^*$J000703.6+155423  & 0751-52251-0577 & 0.1141 & T13 & 41.94  & 97$\pm$45&97$\pm$45\\
$^*$J111847.0+075419   & 1617-53112-0469 & 0.1269 & G11 & 41.49& 25$\pm$2&25$\pm$2\\
J165939.7+183436 & 1567-53172-0113 & 0.1708 & T13 & 42.28 &  69$\pm$3&69$\pm$3\\
J150719.9+002905  & 0310-51616-0565 & 0.1820 & G11 & 42.63&  55$\pm$8&55$\pm$8\\
J075820.9+392335 & 0544-52201-0464 & 0.2162 & G11 & 42.66 &  89$\pm$7&89$\pm$7\\
$^*$J090036.8+205340    & 2282-53683-0103 & 0.2357 & G11 & 42.57  & 55$\pm$2&55$\pm$2\\
  J083454.9+553421  & 5153-56577-0372 & 0.2415 & G11 & 42.34& 53$\pm$5&53$\pm$5\\
J131403.5+365438 & 2032-53815-0204 & 0.2967 & T13 &41.82 & 10$\pm$1&10$\pm$1\\
J134242.3+494439 & 1669-53433-0215 & 0.3473 & T13 &42.68 & 36$\pm$3& 36$\pm$3\\
J153641.6+543505 & 0616-52374-0442 & 0.4471 & G11 & 42.48 &  15$\pm$4&15$\pm$4\\  
J080754.5+494627 & 1780-53090-0094 &0.5753 & W12 & 43.34 & 32$\pm$3& 32$\pm$3\\
\toprule
\end{tabular}
\tablefoot{Column (1) to (5): see Table A.2 for description. Column (6) and (7): NC and OC doubly ionized oxygen flux ratios.
Horizontal lines separate the sub-samples defined in  Sect. \ref{temperaturesection}: the first 8 rows are associated with sources with evidence of ionized outflows and well detected (S/N$>10$) [O III]$\lambda$4363 lines. For these sources, both NC and OC flux ratios have been measured. The following 10 AGNs are not associated with ionized outflows and only NC flux ratios are reported. Finally, the last 26 rows are associated with AGNs with evidence of outflows but faint (5 $<$ S/N $<$ 10) [O III]$\lambda$4363 lines. For these sources, NC and OC flux ratios have been imposed to be equal (i.e., $T_e(NC)=T_e(OC)$). The sources marked with $^*$ are also present in the sample of 28 AGNs with measured [S II] flux ratios. }
\end{table*}

\begin{figure*}[h]
\includegraphics[width=18cm,height=22cm, trim=0 130 30 80,clip]{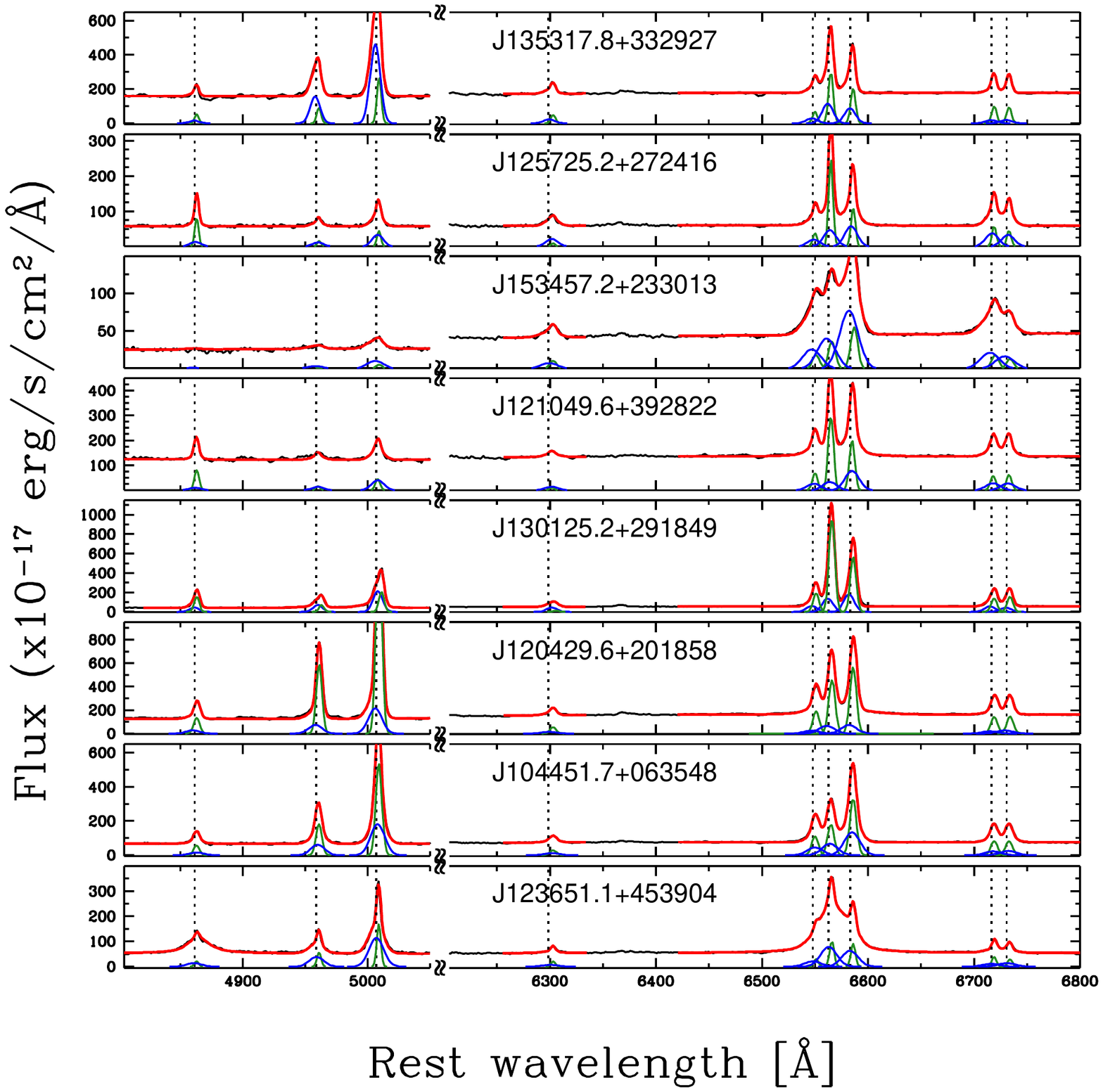}
\caption{Rest-frame optical spectra of 8 out 28 AGNs for which we derive density-sensitive flux ratios. For each object, we show the spectrum (black curve) and the best-fit models (red curves) obtained from multicomponent simultaneous fit in the regions around the H$\beta$ and the H$\alpha$ emission. The dashed vertical lines mark the location of H$\beta$, [O III] doublet, [O I],  H$\alpha$, [N II] and [S II] doublets. Best-fit NC and BC profiles are highlighted with green curves; blue curves display OC emission. }
\label{fit}
\end{figure*}

\begin{figure*}[h]
\ContinuedFloat
\includegraphics[width=18cm,height=24cm, trim=0 130 30 80,clip]{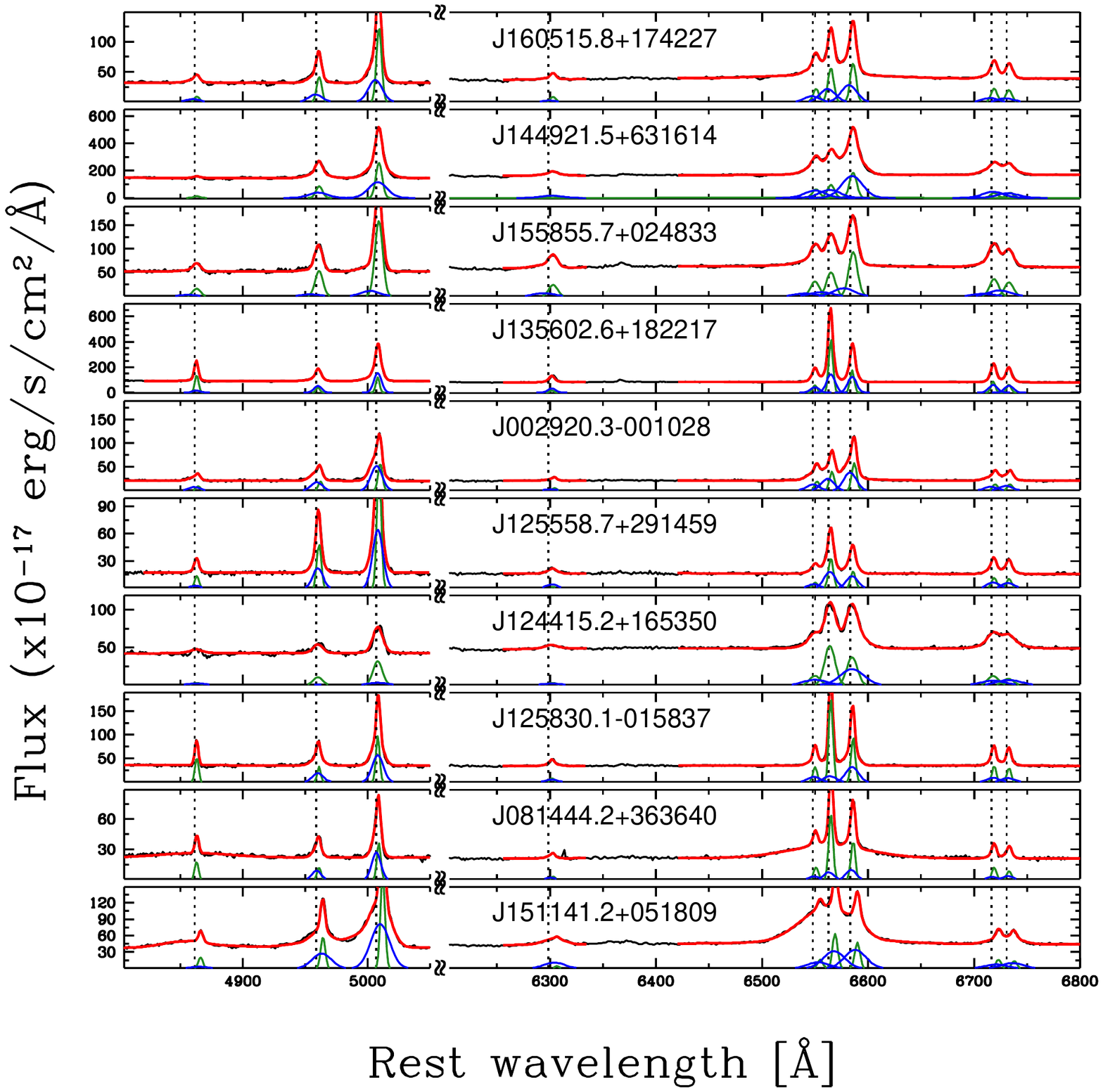} 
\caption{ See the first figure for expansive description. }
\end{figure*}

\begin{figure*}[h]
\ContinuedFloat
\includegraphics[width=18cm,height=24cm, trim=0 130 30 80,clip]{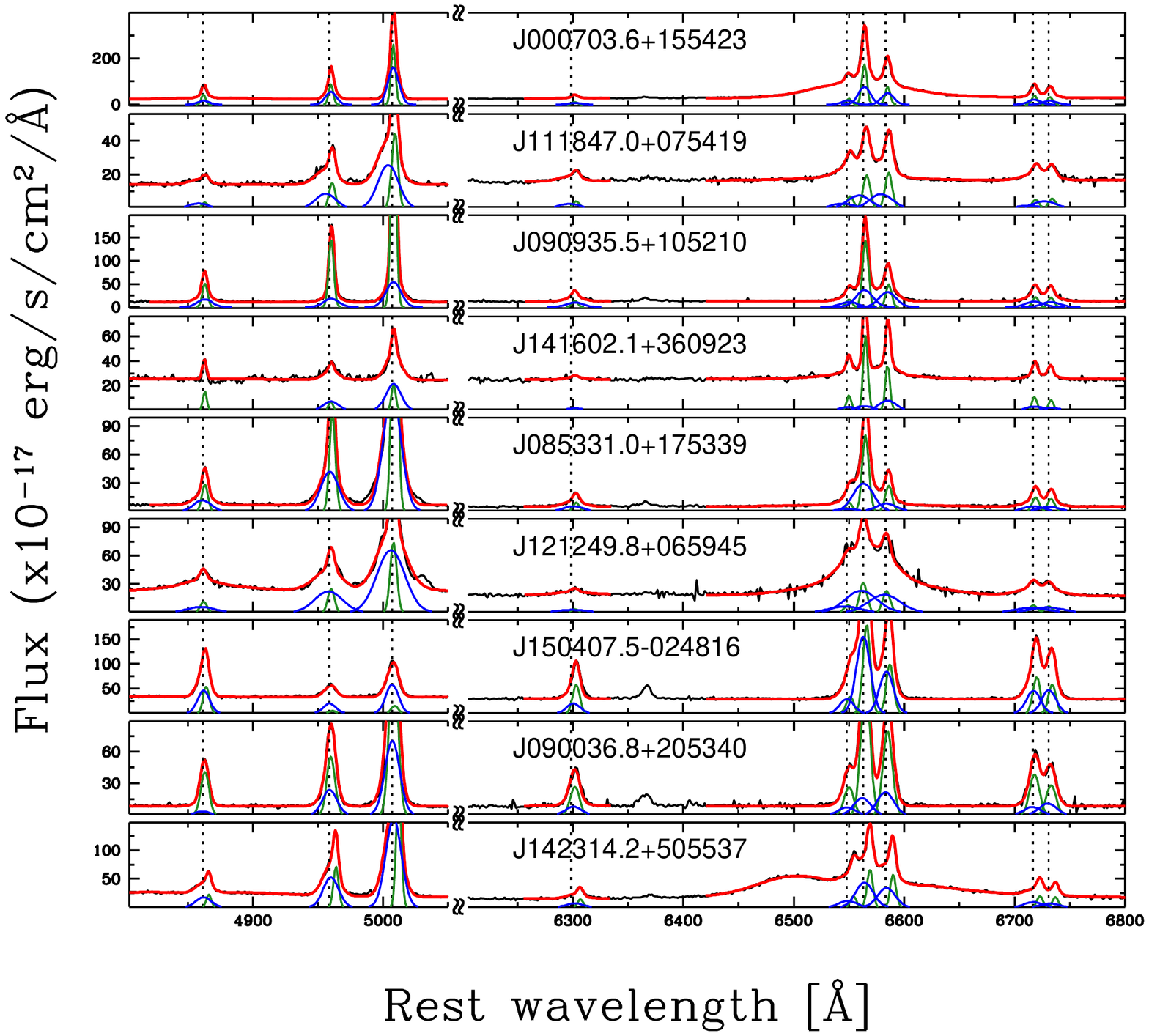} 
\caption{ See the first figure for expansive description. }
\end{figure*}
\end{appendix}

\end{document}